# Dark-exciton driven energy funneling into dielectric inhomogeneities in two-dimensional semiconductors

Haowen Su[a], Ding Xu[a], Shan-Wen Cheng[a], Baichang Li[b], Song Liu[b], Kenji Watanabe[c], Takashi Taniguchi[c], Timothy C. Berkelbach[b,d], James Hone[b], Milan Delor[a*]
a. Department of Chemistry, Columbia University, New York, NY, USA
b. Department of Mechanical Engineering, Columbia University, New York, NY, USA
c. National Institute for Materials Science, Tsukuba, Japan
d. Center for Computational Quantum Physics, Flatiron Institute, New York, NY, USA

**Abstract:**
The optoelectronic and transport properties of two-dimensional transition metal dichalcogenide semiconductors (2D TMDs) are highly susceptible to external perturbation, enabling precise tailoring of material function through post-synthetic modifications. Here we show that nanoscale inhomogeneities known as nanobubbles can be used for both strain and, less invasively, dielectric tuning of exciton transport in bilayer tungsten disulfide ($WSe_2$). We use ultrasensitive spatiotemporally resolved optical scattering microscopy to directly image exciton transport, revealing that dielectric nanobubbles are surprisingly efficient at funneling and trapping excitons at room temperature, even though the energies of the bright excitons are negligibly affected. Our observations suggest that exciton funneling in dielectric inhomogeneities is driven by momentum-indirect (dark) excitons whose energies are more sensitive to dielectric perturbations than bright excitons. These results reveal a new pathway to control exciton transport in 2D semiconductors with exceptional spatial and energetic precision using dielectric engineering of dark state energetic landscapes.

**Main text:**
Two-dimensional transition metal dichalcogenide semiconductors (2D TMDs) are van der Waals materials that hold great promise for nanoscale optoelectronics thanks to their strong light-matter interactions even at the atomically-thin limit. The optoelectronic properties of 2D TMDs are in large part governed by their Coulomb-bound electron-hole pairs (excitons), with relatively large binding energies of up to hundreds of milli-electronvolts (meV) due to weak out-of-plane dielectric screening.[1–6] Unlike free charges, excitons are charge neutral and are therefore difficult to manipulate with external electric fields in electronic devices.[7–9] Therefore, the transport properties of excitons are largely dictated by random, diffusive motion with no long-range directionality, limiting their use as information and energy carriers. Finding new ways to manipulate exciton transport in 2D TMDs without radically altering other material properties would result in excitonic devices that combine strong light-matter interactions and precise control over energy and information flow in atomically thin materials.

An attractive route to controlling the properties of 2D TMDs is to leverage their extreme sensitivity to extrinsic factors such as strain,[10–21] and dielectric screening by the environment (Figure 1a),[5,22–26] enabling post-synthetic tuning of their optoelectronic and transport properties. For example, tensile strain reduces the optical transition energy of 2D TMDs;[16,18,27,28] localized strain regions thus create energy gradients that can funnel and trap excitons at nanoscale low-energy sites, a process that was leveraged to create long-lived quantum emitters.[14,29–33] Strain engineering, however, is difficult to control over macroscopic scales and can introduce undesired disorder.

A less invasive and in principle more controlled approach is to modulate Coulomb interactions in 2D TMDs through dielectric engineering of the surrounding environment, for example by modifying the substrate on which TMD layers are deposited.[22] The dielectric contrast at the interface of monolayer or few-layer TMDs leads to extreme sensitivity of their Coulomb interactions to the environment.[2,6,22,34,35] Although dielectric engineering has indeed been used to successfully modulate some TMD properties, changes in absorption and photoluminescence energies (the optical gap $E_{opt}$) of 2D TMDs as a function of substrate dielectric constant are predicted to be surprisingly small: ~100 meV for a 20-fold increase in substrate dielectric constant.[36] These modest changes arise from a cancellation effect between a large dielectric-induced change of the single-particle bandgap ($\Delta E_g$, the minimum gap between the conduction and valence bands without electron-hole correlations) and a similarly large change in the exciton binding energy ($\Delta E_b$),[36] which overall results in small changes to $\Delta E_{\text{opt}} = \Delta E_g - \Delta E_b$ (using positive binding energies as convention), as depicted in Figure 1b.

Although this weak dependence of $E_{\text{opt}}$ on the dielectric environment limits the applicability of dielectric engineering for bright, momentum-direct (KK) excitons, the near cancellation of two large contributions suggests that small differences in either $\Delta E_g$ or $\Delta E_b$ could substantially enhance the environmental sensitivity of excitons. For example, 2D TMDs sustain a wealth of intervalley, momentum-indirect 'dark' excitons (e.g. KQ and ΓQ excitons, illustrated in Figure 1a for the $K$ and $Q$ valleys) for which the effect of dielectric engineering remains unexplored, largely because dark excitons cannot be directly probed by absorption or photoluminescence at room temperature.[37–43]

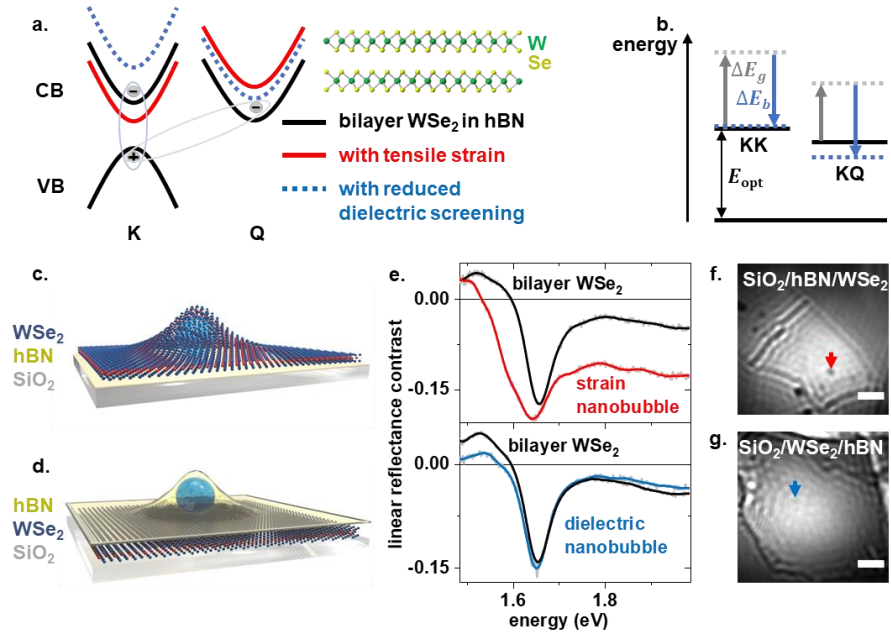

Figure 1. (a) Schematic single-particle valence and conduction bands (energy vs. momentum) for bilayer WSe$_2$ at the $K$ and $Q$ (often labeled Λ) valleys, showing bright KK momentum-direct excitons and dark KQ momentum-indirect excitons. Spin-forbidden dark states are ignored. The proposed effects of tensile strain and reduced dielectric screening on the single-particle bands are illustrated in red and blue, respectively. (b) Exciton energy-level diagram displaying the effect of reduced dielectric screening on the single-particle bandgap ($\Delta E_g$) and binding energy ($\Delta E_b$) in grey and blue arrows, respectively. (c, d) Illustrations of a strain nanobubble (c) and a dielectric nanobubble (d). (e) Linear reflectance contrast, ($R_{\text{sample}}$-$R_{\text{substrate}}$)/$R_{\text{substrate}}$ obtained with an oil-immersion objective for flat regions of bilayer WSe$_2$ (black), strain nanobubbles (red) and dielectric nanobubbles (blue). (f, g) iSCAT images taken at a probe wavelength of 1.62 eV for samples containing strain nanobubbles (f) and dielectric nanobubbles (g). The arrows point to two example nanobubbles; strain nanobubbles are clearly visible, whereas dielectric nanobubbles are barely distinguishable. Scale bars are 2 μm.

In this work, we show that dark excitons exhibit a greater sensitivity to their dielectric environment, with reduced energies in the presence of reduced dielectric screening (Figure 1b), enabling dielectrically engineered exciton transport. We use spatiotemporally resolved optical scattering microscopy[44] to directly image the transport of the total population of excitons in heterogeneous dielectric environments at room temperature. We study bilayer WSe$_2$ because it hosts a number of momentum-dark excitons that lie tens to hundreds of meV lower in energy than the bright KK exciton, ensuring that the majority of excitons populate momentum-indirect states at room temperature.[32,38,40,45] We reveal that excitons can be efficiently funneled into nanoscale dielectric inhomogeneities even though the energies of bright excitons are negligibly affected, from which we draw conclusions about the environmental sensitivity of dark excitons. Leveraging the important contribution of dark excitons and their large dependence on environmental dielectric represents a new approach for highly effective, non-invasive manipulation of exciton transport and dynamics in TMDs.

To manipulate excitons and concentrate them into spatially well-defined regions at room temperature, we use bilayer 2H-$WSe_2$/hexagonal boron nitride (hBN) heterostructures containing nanobubbles. Nanobubbles are nanoscale regions of trapped adsorbates that form naturally between mechanically stacked van der Waals layers in heterostructures[46–49]. In our heterostructures, the nanobubbles create regions of strain either in $WSe_2$ or in hBN, as illustrated in Figures 1c and 1d, respectively. We control the type of nanobubbles formed through the stacking order (Supplementary Section 1). 'Strain nanobubbles' occur when the $WSe_2$ is strained, leading to the well-known strain-induced modulation of exciton energies[49,50] (Figure 1a). In contrast, the nanobubbles formed in hBN create nanoscale regions of slightly lower dielectric screening wherever the hBN and $WSe_2$ are not in contact, without directly affecting the $WSe_2$. We term these regions 'dielectric nanobubbles'. Our structures therefore allow direct comparison between the effect of strain and dielectric inhomogeneities on exciton transport. The bilayers are mechanically exfoliated from ultra-pure flux-grown crystals[51,52] that minimize the influence of intrinsic defects on exciton dynamics, providing confidence in our assignment of strain and dielectric effects on exciton transport.

Figure 1e shows the linear, room-temperature reflectance contrast spectra near the A-exciton (bright KK) resonance for $WSe_2$ at strain and dielectric nanobubbles compared to nearby flat regions. The strain nanobubble exhibits a broader spectrum and a redshift of ~ 65 meV compared to the flat region. In contrast, the $WSe_2$ spectrum changes negligibly at the location of dielectric nanobubbles, suggesting that single-side hBN encapsulation of glass-supported bilayer $WSe_2$ barely influences the optical gap. Both of these observations are consistent with previous reports.[22,49] In Figure 1f,g, we display interferometric scattering microscopy (iSCAT) images of the two types of heterostructures. iSCAT is an extremely sensitive version of brightfield reflectance microscopy.[53–55] Strain nanobubbles are clearly visible in iSCAT, similarly to brightfield microscopy; dielectric nanobubbles, however, are barely distinguishable in iSCAT, and invisible in brightfield microscopy. As such, although nanobubbles form regularly in van der Waals heterostructures (as observed for example in AFM images, *vide infra*), they can be easily overlooked if they only strain transparent screening layers.

To determine the effect of nanobubbles on exciton transport, we turn to a pump-probe version of iSCAT, stroboscopic scattering microscopy (stroboSCAT).[44] stroboSCAT is a spatiotemporally resolved far-field imaging approach[56,57] wherein a diffraction-limited optical excitation generates a population of excitons, and a widefield backscattering optical probe images the spatial distribution of photo-generated excitons at controllable time delays later (Figure S2). These measurements directly track the motion of excitons in real space with picosecond resolution and few-nanometer precision. Contrast in stroboSCAT is generated by pump-induced changes to material polarizability,[44,58,59] and is thus sensitive to species such as momentum-dark excitons because their presence modifies the polarizability of optically-allowed transitions, even if there is no direct population exchange between dark and bright excitons. In contrast, photoluminescence (PL), a powerful contrast mechanism often used for spatiotemporal imaging of exciton transport in TMDs,[60–62] requires population transfer from dark to bright states followed by recombination through momentum-allowed channels, or alternative brightening mechanisms, to be sensitive to momentum-dark excitons at room temperature. We show below that these differences in contrast generation give rise to important discrepancies between stroboSCAT and PL images, and uniquely enable stroboSCAT to monitor momentum-dark excitons trapped in deep potential wells. To perform the experiments described below in the linear regime (without exciton trap saturation), pump excitation fluences are limited to the generation of ~10 excitons per diffraction-limited spot. We achieve the necessary sensitivity for probing such low exciton densities by tuning our probe energy slightly lower than the optical gap, operating in a pre-resonance regime that dramatically enhances backscattering contrast without perturbing the sample (Supplementary Section 2).

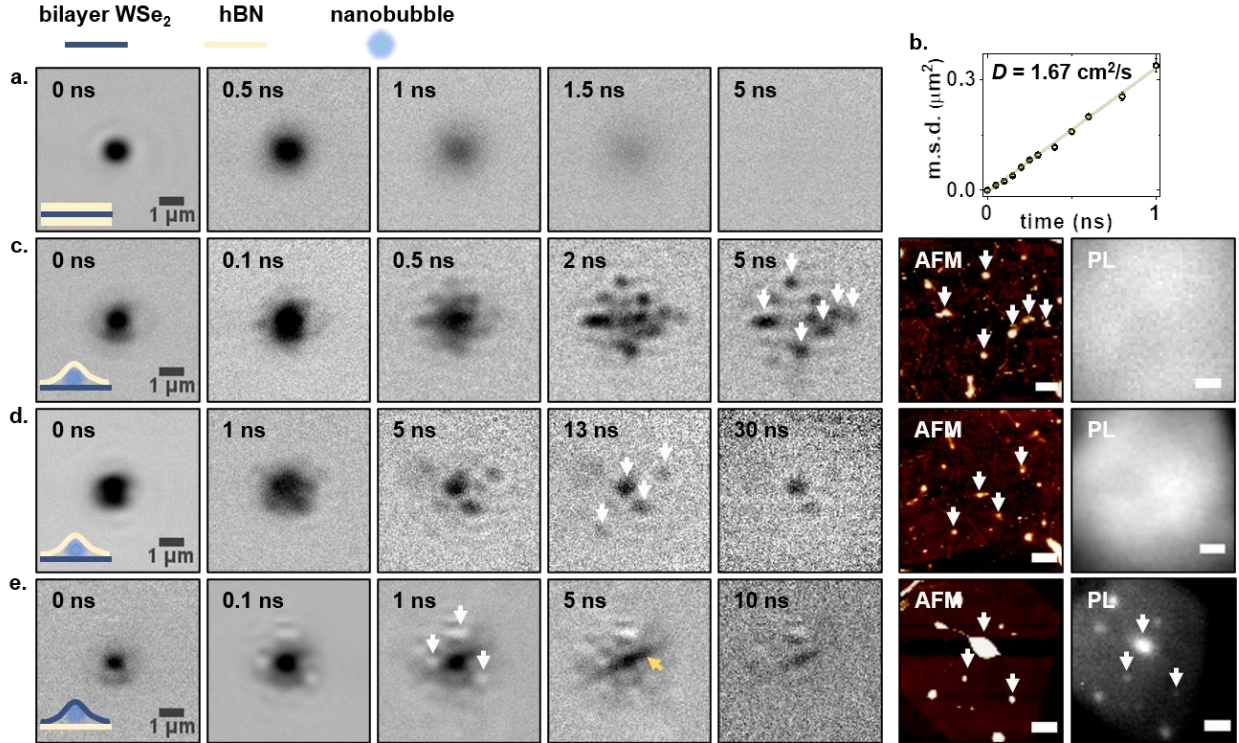

Figure 2. (a) stroboSCAT images of exciton transport in fully-encapsulated bilayer WSe$_2$. (b) Mean squared displacement (m.s.d.) for the data shown in panel (a). The line is a linear fit to the data (circles), providing a diffusion coefficient $D$ = 1.67 cm$^2$/s for excitons in encapsulated bilayer WSe$_2$. (c, d) stroboSCAT, AFM and PL images for bilayer WSe$_2$/hBN heterostructures in regions with dielectric nanobubbles. We observe clear exciton localization with stroboSCAT, at locations that correlate with AFM images of the nanobubbles, as indicated by white arrows; PL images (taken with widefield illumination using a 1.9 eV laser), however, do not show any evidence of exciton localization. (e) stroboSCAT, AFM and PL images for a bilayer WSe$_2$/hBN heterostructure in a region with strain nanobubbles. Excitons localizing in strain nanobubbles switch to bright contrast, and correlate with both AFM and PL images. Additional features (e.g. gold arrow) indicate a disordered energetic landscape, as detailed in the text. For all stroboSCAT datasets, pump and probe wavelengths are 2.82 eV and 1.62 eV, respectively. The pump fluence is ~150 nJ/cm$^2$, corresponding to ~10 excitons per diffraction-limited spot which avoids trap saturation effects that occur at higher pump fluences. All scale bars are 1 μm.

Figure 2a displays stroboSCAT images taken on bilayer WSe$_2$ encapsulated on both sides by hBN. The expanding dark contrast corresponds to diffusing pump-generated excitons. Full hBN encapsulation is known to considerably reduce the effect of dielectric disorder on exciton dynamics.[34,63–65] Indeed, our measurements on fully encapsulated WSe$_2$ indicate that even if nanobubbles are present in the top hBN layer, effective dielectric screening is still provided by the bottom layer. As a result, the exciton dynamics in WSe$_2$ are virtually oblivious to nanobubbles, in stark contrast to the half-encapsulated samples described below. Thus, in fully encapsulated samples, we observe perfectly isotropic and diffusive exciton transport. From the mean squared displacement (m.s.d.) of the exciton population, $\sigma^2(t) - \sigma^2(0) = 2Dt$ (Supplementary Section 1.3), where $\sigma(t)$ is the Gaussian width of the exciton spatial distribution at pump-probe time delay $t$, we extract an intrinsic diffusivity $D = 1.67 \pm 0.03$ cm$^2$/s (Figure 2b) for excitons in encapsulated bilayer WSe$_2$.

Our key results are displayed in Figures 2c,d, which show two examples of exciton dynamics in single-side encapsulated samples in the presence of dielectric nanobubbles. We observe visually striking evidence of excitons localizing at nanoscale sites on nanosecond timescales, after which the excitons stop migrating and remain in some cases for over 30 nanoseconds. We confirm that the trap locations (highlighted with white arrows) are nanobubbles by spatially correlating the stroboSCAT images to atomic force microscopy images (AFM, middle-right panel). The AFM images show ~5-30 nm tall topographical features that are characteristic of nanobubbles. The locations of these nanobubbles correspond precisely to the exciton localization identified in stroboSCAT. Conversely, PL images (right panels) exhibit subtle micron-scale inhomogeneities, as observed in other studies of dielectric disorder,[34] but remarkably do not display any evidence of the presence of nanobubbles. This lack of correspondence between PL images and stroboSCAT is a strong indication that the observed exciton dynamics in bilayer $WSe_2$ are not governed by bright, emissive states.

To further elucidate the role of momentum-dark excitons in these surprising observations, we performed stroboSCAT measurements on monolayer $WSe_2$/hBN heterostructures, in which bright states form a substantial fraction of total exciton population at room temperature.[38,45,66] Our results indicate that dielectric nanobubbles only subtly perturb exciton transport dynamics in monolayer $WSe_2$, and we find no evidence of long-lived exciton localization in the nanobubbles (Figure S4). These results lend further support to our hypothesis that momentum-indirect states govern exciton dynamics in bilayer $WSe_2$, and are responsible for the dramatic modification of exciton transport properties in the presence of dielectric inhomogeneities.

To compare the effects of dielectric engineering with the more commonly-encountered strain engineering, we now turn to bilayer $WSe_2$/hBN heterostructures containing strain nanobubbles (Figure 2e). Tensile strain in the nanobubbles causes the optical gap of $WSe_2$ to be lowered by several tens of meV compared to surrounding unstrained regions (Figure 1e), and raises the KQ exciton energies by a similar amount (Figure 1a).[32,38,50,67,68] We observe exciton localization in these nanobubbles (indicated by white arrows, spatially-correlated to AFM images), but with several important differences compared to the dielectric nanobubbles of Figures 2c,d. First, the stroboSCAT signal for the trapped excitons in strain nanobubbles is inverted, showing bright rather than dark contrast. This behavior is expected in pre-resonant stroboSCAT when probing optically-accessible lower-energy sites (Supplementary Section 2). Second, we observe significantly enhanced PL at these nanobubbles (Figure 2e, right panel), in agreement with previous reports.[48,49] Together, these results confirm that exciton dynamics within strain nanobubbles in bilayer $WSe_2$ are governed by bright KK excitons, which can lie lower in energy than KQ excitons under sufficient tensile strain.[38] The clear differences observed between dielectric and strain nanobubbles also confirm that the modulation of exciton dynamics in dielectric nanobubbles occurs through a different mechanism than semiconductor strain.

Finally, the stroboSCAT images of Figure 2e indicate a strongly disordered landscape, with localization and long-lived states appearing not just in nanobubbles but also in other inhomogeneities not observed in the PL image. For example, the gold arrow in Figure 2e highlights a long-lived, line-like feature with dark contrast appearing between nanobubbles. This feature may arise from regions of compressive strain between closely spaced nanobubbles, which carries an opposite effect to tensile strain (raised KK and lowered KQ exciton energies)[38,67,68] and could lead to dark-state trapping that would explain the absence of this feature in the PL image. Similarly, recent work suggests that momentum-dark excitons experience anti-funneling (*i.e.* move away) from tensile strain regions,[69] which may explain our observation of trapped momentum-indirect excitons between closely-spaced strain nanobubbles.

To provide further insight into how dielectric nanobubbles modify exciton transport, Figure 3a displays the dynamics of excitons taken from the data in Figure 2d, demonstrating that the lifetime of excitons trapped in dielectric nanobubbles is dramatically prolonged. For example, for the nanobubble at the top-right of the excitation spot in Figure 2d, we extract a lifetime of 12 nanoseconds (blue curve in Figure 3a), a 26-fold increase over the lifetime of excitons in flat regions of the bilayer $WSe_2$/hBN heterostructures. Assuming the lifetime is dictated by thermally activated de-trapping from the low-energy sites in the nanobubble, and ignoring entropic contributions, a 26-fold lifetime enhancement suggests that an 84 meV-deep trap is formed by dielectric nanobubbles. Strain nanobubbles lead to more modest exciton lifetime enhancements (Figure S5) in our heterostructures, an unsurprising finding since the bright states in strain nanobubbles relax through momentum-allowed channels, in contrast to the de-trapping and intervalley scattering required for the relaxation of momentum-dark excitons trapped in dielectric nanobubbles.

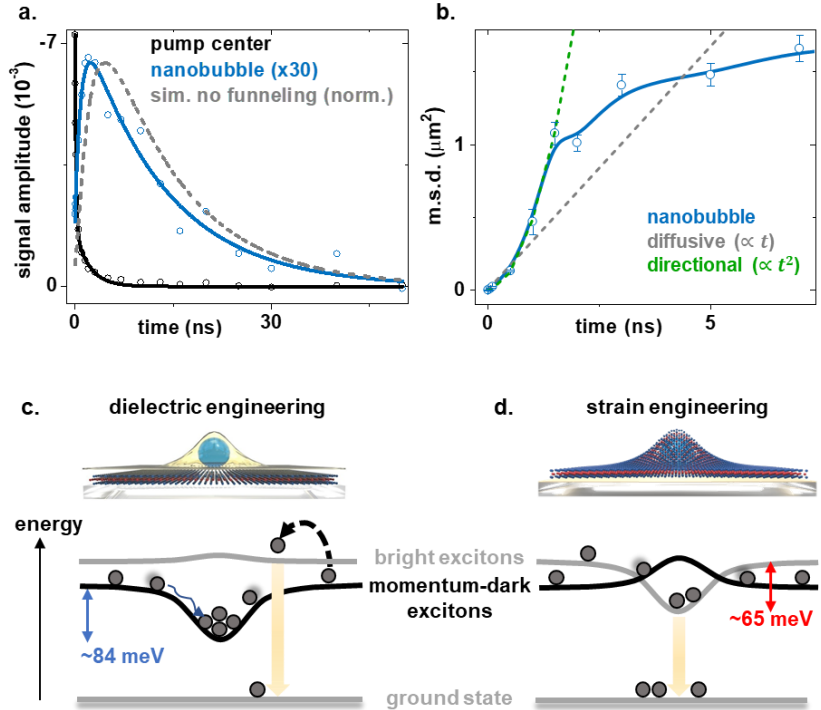

Figure 3. (a) Exciton dynamics in dielectric nanobubbles (from stroboSCAT data in Figure 2d). The lines are biexponential fits to the data that we use to extract the lifetime of trapped excitons. The grey trace is calculated from an isotropic diffusion model with $D = 1.67$ cm$^2$/s, showing that excitons funnel into nanobubbles faster than expected from a purely diffusive model. (b) Mean squared displacement obtained for exciton transport towards a nanobubble for the data in Figure S6 and Supplementary Movie 1 (blue data in open circles, with a spline curve to guide the eye). The m.s.d. toward the nanobubble exceeds that expected for purely diffusive transport with $D = 1.67$ cm$^2$/s (grey trace). A superdiffusive fit (green trace) to the early-time data suggests the m.s.d. is initially proportional to $t^2$, indicating funneling into nanobubbles. (c, d) Summary energy-level diagrams illustrating exciton dynamics in bilayer WSe$_2$/hBN heterostructures in the presence of dielectric (c) and strain (d) nanobubbles. The quoted trapping energies are representative numbers extracted from datasets presented in the text, but can vary from site to site. Black dashed arrows indicate thermally-activated de-trapping or intervalley scattering, while yellow arrows indicate radiative recombination.

Comparing the observed rise-time of the signal at dielectric nanobubbles with the expected rise time calculated from the known diffusivity of 1.67 cm$^2$/s (grey trace in Figure 3a), we find that excitons reach the nanobubbles more than twice as fast as would be expected in a purely diffusive model (time of maximum population is 2.2 ns instead of the expected 4.7 ns). These results indicate that downhill energy gradients caused by dielectric nanobubbles cause directional funneling of excitons toward the nanobubbles. Exciton funneling is visually most evident in stroboSCAT movies of exciton migration in samples with just a few nanobubbles (Supplementary Movies 1 and 2). Figure 3b plots the m.s.d. for exciton transport into a nanobubble from the data in Supplementary Movie 1 (analysis in Figure S6). For purely random, diffusive behavior, we expect the m.s.d. to be proportional to time, as indicated in the grey dashed trace in Figure 3b for $D = 1.67$ cm$^2$/s. Instead, exciton transport towards the nanobubble is superdiffusive: We extract an

m.s.d. proportional to $t^2$ prior to the excitons trapping in the nanobubble (the green trace in Figure 3b is a superdiffusive fit to the first five blue data points). This behavior indicates directional transport (drift) of excitons into the nanobubble due to a progressive downhill energy gradient created by the dielectric perturbation of the nanobubble. Our data implies a drift velocity of 640 nm/ns sustained over a distance of 1 μm, similar to the exciton drift velocities and distances achieved using large strain gradients in monolayer TMDs[13,62]. These results provide additional confirmation that excitons are indeed efficiently funneled, rather than simply trapped, by dielectric inhomogeneities, providing a pathway to manipulating exciton transport with high precision over macroscopic scales, and to concentrate excitons in target regions with nanoscale precision.

Overall, our results indicate that dielectric engineering is particularly effective for TMDs in which the majority of excitons reside in low-lying momentum-indirect states. Leveraging this effect, we show that nanobubbles in a hBN dielectric screening layer provides an efficient path to controlling exciton dynamics in bilayer $WSe_2$. As summarized in Figures 3c,d, the use of nanobubbles allows a direct comparison of tensile strain and dielectric engineering. Tensile strain allows funneling of bright excitons into bright states, advantageous for emitter-based functions. Dielectric engineering, however, is less invasive and achieves longer exciton lifetimes through momentum-dark excitons, valuable for light harvesting and information processing applications. Our experimentally observed dependence of momentum-indirect excitons on their dielectric environment is consistent with their larger effective mass and therefore larger binding energy renormalization than KK excitons,[37–39,70] combined with a reduced sensitivity in their band gap. The latter has been observed in a recent GW calculation of monolayer $MoS_2$,[71] though it was not observed in a similar previous calculation on monolayer $WS_2$.[72] Resolving these small energy differences due to dielectric effects is an ongoing challenge for theoretical calculations.

We anticipate that patterning of high-dielectric substrates such as graphite[22] on which bilayer TMDs are deposited will provide a platform to predictively manipulate exciton transport with high precision for new, dark-exciton driven functionalities in nanoscale devices. The combination of efficient funneling and long lifetimes of momentum-dark excitons in dielectric inhomogeneities could be leveraged to drive spatially-confined population inversion, as well as to concentrate excitons generated across macroscopic areas into nanoscale catalytically-active regions to drive photocatalysis under low-light conditions, a strategy similar to exciton funneling from light harvesting antennae into reaction centers in photosynthetic organisms.

**Supporting Information**. Materials and Methods (Figures S1-S2), details of pre-resonant stroboSCAT contrast mechanism (Figure S3), and additional data (Figures S4-S6, Supplementary Movies S1-S2).

**Corresponding Author.** *Milan Delor (milan.delor@columbia.edu)

**Acknowledgments**

We thank Thomas P. Darlington and Professor P. James Schuck (Columbia University) for helpful discussions. This material is based upon work supported by the National Science Foundation under Grant No. DMR-2115625. Synthesis of $WSe_2$ (S.L., J.H.) was supported by the NSF MRSEC program through Columbia in the Center for Precision-Assembled Quantum Materials (DMR-2011738). Synthesis of hBN (K.W. and T.T.) was supported by the Elemental Strategy Initiative conducted by the MEXT, Japan (Grant JPMXP0112101001), and JSPS KAKENHI (Grant Numbers JP19H05790 and JP20H00354). The Flatiron Institute is a division of the Simons Foundation.

# Supplementary information for

# Dark-exciton driven energy funneling into dielectric inhomogeneities in two-dimensional semiconductors

Haowen Su[a], Ding Xu[a], Shan-Wen Cheng[a], Baichang Li[b], Song Liu[b], Kenji Watanabe[c], Takashi Taniguchi[c], Timothy C. Berkelbach[b,d], James Hone[b], Milan Delor[a*]

a. Department of Chemistry, Columbia University, New York, NY, USA
b. Department of Mechanical Engineering, Columbia University, New York, NY, USA
c. National Institute for Materials Science, Tsukuba, Japan
d. Center for Computational Quantum Physics, Flatiron Institute, New York, NY, USA

## 1. Materials and Methods

**1.1 Assembly of WSe$_2$/hexagonal Boron Nitride (hBN) heterostructures:** Bilayer WSe$_2$ flakes are obtained on Si/(100nm) SiO$_2$ wafers using mechanical exfoliation from self-flux-grown single crystal WSe$_2$. The flakes of interest were first identified by color contrast under an optical microscope (Figure S1a), and then characterized by AFM height profile (Figure S1b).[1] We use the dry transfer technique to create the WSe$_2$/hBN heterostructures containing nanobubbles.[2]

Depending on the stacking order, two types of nanobubbles are formed. Specifically, for dielectric nanobubbles, a few-layer hBN flake is picked up by a polypropylene carbonate (PPC) coated polydimethylsiloxane (PDMS) cube. This hBN is then brought into contact with the WSe$_2$ flake of interest. At a temperature between 40°C to 55°C, the adhesion forces between hBN and bilayer WSe$_2$ are strong enough to pick up the bilayer WSe$_2$ flake from the wafer surface. Nanobubbles between the hBN layer and WSe$_2$ form during the adhesion process. Throughout the adhesion process, WSe$_2$ remains on the flat wafer surface that precludes the nanobubble from straining the WSe$_2$, forming instead in the hBN. After picking up the WSe$_2$ with the hBN, the heterostructure is released onto a pre-cleaned glass substrate at a temperature around 100 °C.

In contrast, for strain-induced nanobubbles, we first transfer a few-layer hBN flake onto a pre-cleaned glass substrate using PPC with the method mentioned above. We then pick up the WSe$_2$ flake using a polycaprolactone (PCL) stamp, and stack the WSe$_2$ flake on top of the hBN flake at a temperature between 52°C to 56°C for stronger adhesion. In this case, nanobubbles are formed when releasing the bilayer WSe$_2$ flake on top of the hBN, preferentially straining the WSe$_2$ flake and therefore forming strain nanobubbles.

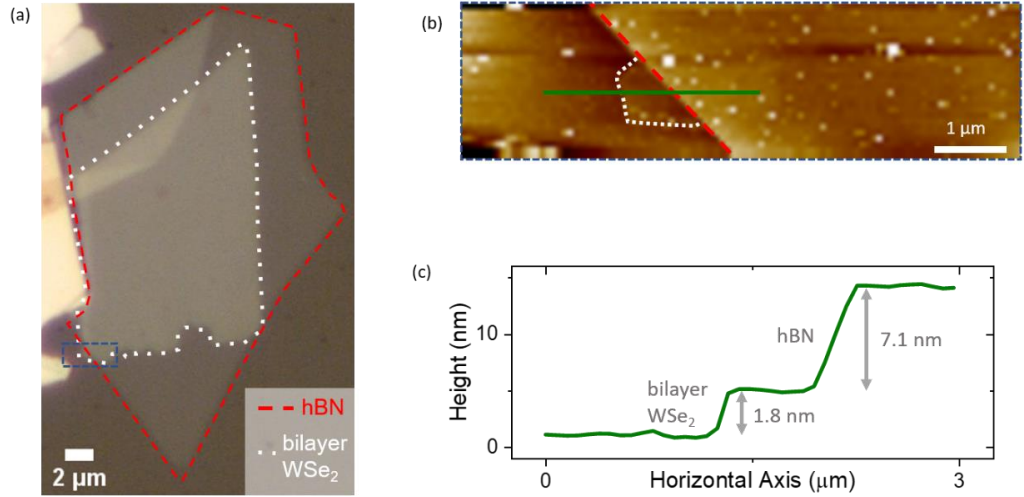

Figure S1. **WSe$_2$/hBN heterostructure characterization**. (a) Optical image of WSe$_2$ heterostructure with hBN layer on the bottom (against glass) and bilayer WSe$_2$ on top. (b) Zoomed-in region of the blue box in (a). The dotted white line marks the boundary of bilayer WSe$_2$ region and the dotted red line distinguishes the bare WSe$_2$ from the WSe$_2$/hBN heterostructure. (c) AFM height profile of the green line section. Showing the bilayer WSe$_2$ thickness of 1.8 nm and hBN thickness of 7.1 nm.

## 1.2. Optical measurements and analysis

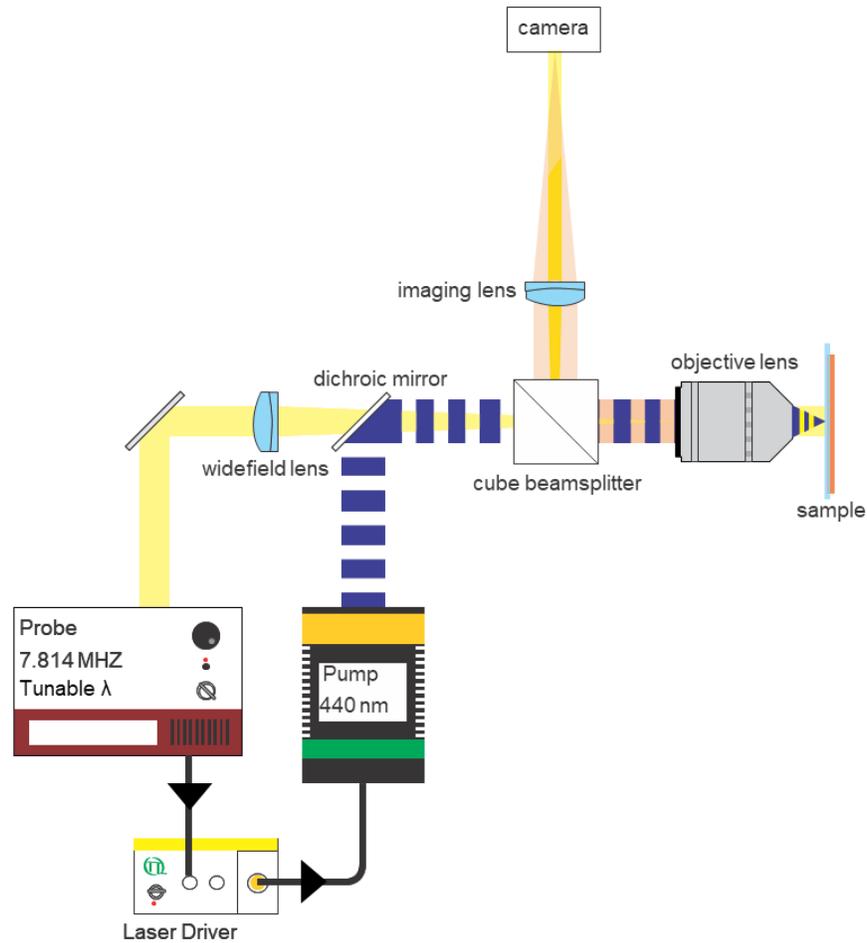

Figure S2. Schematic of pre-resonant stroboSCAT using a laser diode pump and a tunable supercontinuum laser probe.

**Pre-resonant stroboscopic scattering microscopy (stroboSCAT):** stroboSCAT measurements are performed in a home-built microscope (Figure S2) of similar design to previous implementations,[3] but with modifications to the probe laser source allowing precise tuning of the probe wavelength into the pre-resonance regime. For all data shown in the text, the probe pulses are obtained from a NKT Photonics SuperK Extreme white light supercontinuum laser coupled into an acousto-optic tunable filter to select a wavelength of 765 ± 4 nm (15-20 nm below resonance) with a ~30 ps pulsewidth. The pump pulses at 440 nm (60 ps pulsewidth) are obtained from a PicoQuant laser diode (LDH-D-C-440) driven by a PicoQuant laser driver (PDL-828-S "SEPIA II" equipped a SOM 828-D oscillator). The two laser sources are synchronized by triggering the laser diode driver using a pulse train synchronization signal from the supercontinuum laser. All experiments are conducted at room temperature and at a repetition rate of 7.814 MHz. Pump and probe pulses are linearly polarized and parallel to one another. Pump-probe time delays are controlled using the electronic delay capabilities of the diode driver, which are used to negatively delay the diode pulses with respect to the supercontinuum pulses with 20 ps resolution.

Both pump and probe beams are spatially-filtered through pinholes before combining at a dichroic mirror. A 50/50 beamsplitter transmits the pump and probe light into an oil-immersive objective (Leica HC Plan

Apo x63, NA 1.40) and onto the sample. A f = 300 mm widefield lens is inserted in the probe beam path upstream of the dichroic mirror to focus the probe in the back focal plane of the microscope objective, resulting in a collimated, widefield illumination on the sample plane. The pump beam (~7 mm diameter) is sent collimated into the objective, resulting in a focused pump pulse on the sample plane. Probe light reflected from the sample-substrate interface and scattered light from the sample are collected along the same path through the objective and are directed to a complementary metal-oxide-semiconductor camera (FLIR BFS-U3-28S5M-C with a Sony IMX 421 global shutter sensor) through an imaging lens (f = 500 mm). A combination of bandpass and long-pass filter is used to remove residual pump light. Reflected and scattered probe light interfere and generate an image at the camera plane.[3]

We acquire pump ON and pump OFF images at 900 Hz. The pump pulse train is modulated by the laser driver at 900 Hz, corresponding to 8680 light pulses per camera frame, to acquire consecutive pump ON and pump OFF camera frames. To generate stroboSCAT images, we divide pump ON frames by consecutive pump OFF frames, typically averaged over 10,000 pairs of images. Camera pixels are binned 2 x 2 and we acquire images over 200 x 200 pixels after binning. The total magnification for the stroboSCAT images collected is 157.5x, corresponding to 57 nm/binned pixel, and a field of view of 11 x 11 microns in our standard configuration.

**stroboSCAT analysis:** For typical, isotropic diffusion (e.g. data in Figure 2a of the main text), stroboSCAT data can be analyzed similarly to other spatiotemporally-resolved microscopies.[4,5] We plot a center-line profile for each time delay along any given axis, integrating across 4 bin-wide rectangular regions. The resulting profile is fit with a Gaussian function for each time delay:

$$y(t) = A(t) * \exp\left(-\frac{(x - x_c)^2}{2\sigma^2(t)}\right)$$

where A(t) is a pre-exponential factor dependent on the contrast magnitude at each time delay $t$, $x_c$ is the center position, and $\sigma(t)$ is the Gaussian standard deviation for each time delay.

As detailed in Akselrod *et al.*[6], using the property that the variance of convolved Gaussians are additive, the solution to the diffusion equation in one dimension can be expressed as:

$$\langle x(t)^2 \rangle = \sigma^2(t) = \sigma^2(0) + 2Dt$$

where $\langle x(t)^2 \rangle$ is the mean square displacement (m.s.d.), $\sigma^2$ is the variance of the population distribution at any given time, and $D$ is the diffusion coefficient. Thus, for isotropic, random, diffusive motion:

$$D = \frac{\sigma^2(t) - \sigma^2(0)}{2t}.$$

In general, $\sigma^2(t) - \sigma^2(0) = 2D_0 t^\alpha$, where $0 < \alpha \leq 2$. When $\alpha > 1$, transport is superdiffusive, which can occur either in systems where exciton-lattice scattering is suppressed, or when transport is directional (e.g. in energy funnels).

In much of the data shown in the text, the inhomogeneous energy landscapes of the samples being studied poses an analysis challenge, where the standard Gaussian-expansion analysis described above is inapplicable. To quantify exciton migration in these landscapes, we instead identify transport pathways toward a specific nanobubble, and fit the center-of-mass position of the signal as it migrates towards the nanobubble using multi-Gaussian fitting. This analysis is performed in Figure S7, for the m.s.d. plotted in Figure 3b of the main text.

**Linear reflectance:** linear reflectance spectra are collected from the same microscope using a fiber-coupled stabilized tungsten-halogen white light source (Thorlabs SLS201L) coupled into the objective. The collected reflection is sent to a prism spectrometer.

2. **Pre-resonant contrast enhancement and stroboSCAT signal sign**

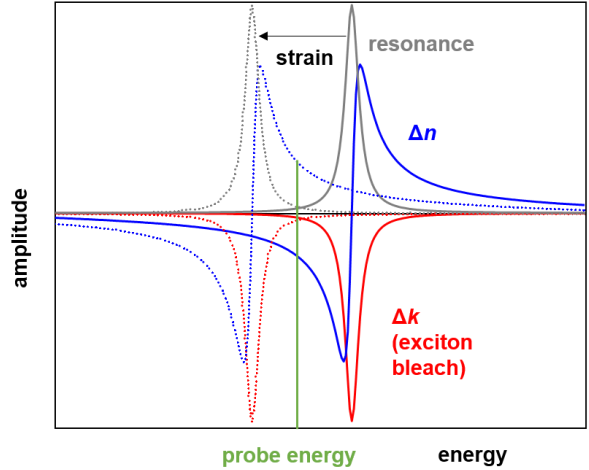

Figure S3. Relationship between exciton resonance (grey trace), pump-induced change in the imaginary part of the refractive index ($\Delta k$, red trace, corresponding to exciton bleaching), and pump-induced change to the real part of the refractive index ($\Delta n$, blue trace), in an idealized scenario with Lorentzian lineshapes and no excited state absorption. In pre-resonant stroboSCAT, the probe energy (represented with the green line) is chosen to maximize $\Delta n$ and minimize $\Delta k$, which is best done a few meV below gap. If strain shifts the energy of the exciton resonance by tens of meVs (represented as the change between solid lines and dotted lines), then $\Delta n$ can switch from negative to positive contrast at the same probe energy.

Contrast in stroboSCAT is primarily generated by pump-induced changes to the real part of the refractive index, $n$. Changes in $n$ occur through both optical polarizability changes, typically due to changes in resonance (e.g. pump-induced absorption bleaching), or through DC polarizability changes (e.g. a pump-induced change in local material density). Changes in absorption ($\Delta k$) are related to $\Delta n$ through a Hilbert transform, as illustrated in Figure S3. In pre-resonant stroboSCAT, we use the fact that $\Delta n$ can be very large near the optical gap in excitonic samples. We choose a below-gap probe energy that maximizes $\Delta n$ (blue trace in Figure S3) while minimizing $\Delta k$ (red trace). This ideal, 'pre-resonant'[7] probe energy allows maximizing stroboSCAT contrast while ensuring that very little probe light is absorbed by the sample. As such, the probe energy can be increased to fill the well depth of the camera without risking sample damage or nonlinear interactions. Filling the camera well depth allows minimizing shot-noise and thus increasing signal-to-noise ratio by a factor of $\sqrt{N}$, where $N$ is the number of detected probe photons in our shot-noise limited measurements.

As a consequence of probing in a pre-resonance configuration, small changes to the optical gap carry large consequences in stroboSCAT signal contrast. As illustrated in Figure S3, when the optical resonance (grey) redshifts, for example due to strain in bilayer $WSe_2$, $\Delta n$ can switch from negative (solid blue line for unstrained) to positive (dotted blue line for strained) at our probe energy. This effect is responsible for the contrast switch in stroboSCAT observed in Figure 2e of the main text.

## 3. Supplementary data

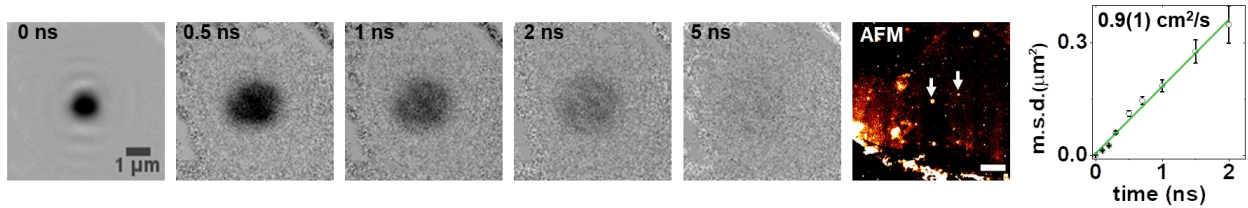

Figure S4. **Exciton transport in monolayer WSe$_2$/hBN heterostructure with dielectric nanobubbles**. stroboSCAT images are captured in a region containing nanobubbles of similar height and radius as the nanobubbles observed in bilayer heterostructures. No obvious trapping effect is observed in the monolayer WSe$_2$/hBN heterostructure, in stark contrast to bilayers. Although some disorder is clearly present in the stroboSCAT images, exciton transport appears almost isotropic and diffusive, with a diffusivity of 0.9 ± 0.1 cm$^2$/s (right panel). We have reproduced these results across multiple regions with nanobubbles of varying height and radius. These results indicate that bright excitons in monolayer WSe$_2$ are only subtly affected by dielectric disorder, in stark contrast to momentum-dark excitons in bilayer WSe$_2$.

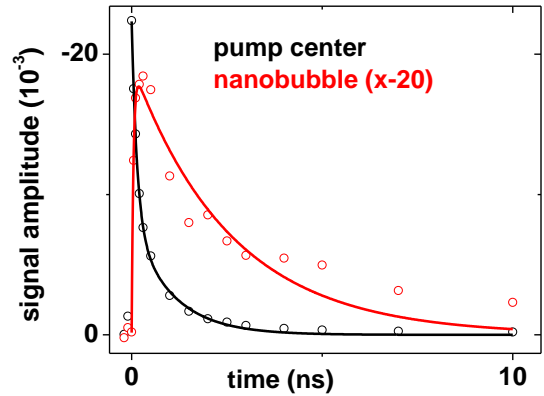

Figure S5. **Exciton dynamics in strain nanobubbles**. Exciton dynamics in a strain nanobubble (red data points, fit with a biexponential growth-decay function). We extract a nanobubble lifetime of 2.6 nanoseconds, an almost six-fold increase over the lifetime of excitons in flat regions of bilayer WSe$_2$. The lifetime of excitons in dielectric nanobubbles is consistently longer than in strain nanobubbles in our samples.

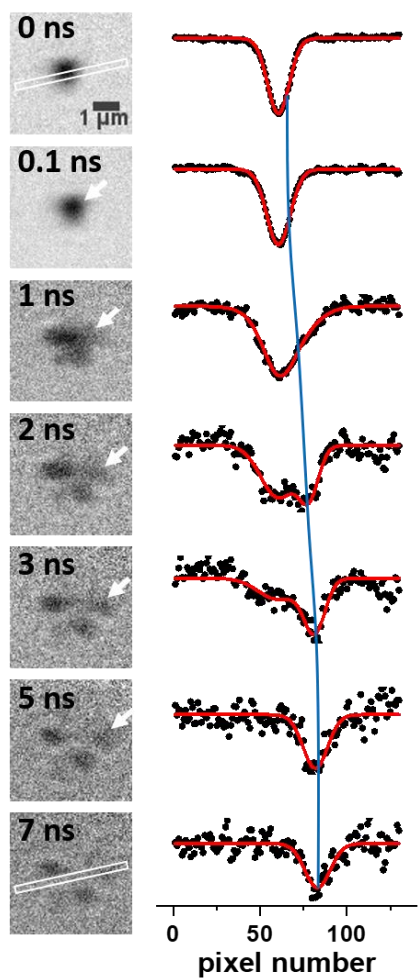

Figure S6. Analysis for m.s.d. curve plotted in Figure 3b of the main text (also see supplementary movie 1). For non-isotropic diffusion, the signal clearly does not remain Gaussian over time. In this example of energy funneling into three nanobubbles close to the excitation spot, stroboSCAT images indicate that the exciton population 'splits' away from the initial excitation region into three distinct spatial distributions that migrate to the nanobubbles before trapping. This behavior poses an analysis challenge as the standard Gaussian expansion analysis cannot be used. Instead, we isolate and track the spatial distribution of the exciton population migrating towards a single nanobubble, as illustrated with the white arrows. To do this, we extract stroboSCAT profiles across the rectangular profile drawn on the images. From 0 to 0.5 ns, i.e. before the exciton population splits, we fit a normal Gaussian expansion as described in supplementary section 1.3 for isotropic diffusion. From 1 to 3 ns, part of the population splits away and migrates towards the nanobubble. Here, we use double-Gaussian fitting, fixing the center of a first Gaussian to the same center position as the image at 0 ns (representing the central, 'unsplit' decaying population), and fitting the position of a second Gaussian corresponding to the exciton population that is migrating towards the nanobubble. The separation between the centers of the second Gaussian and the fixed Gaussian thus corresponds to the distance traveled by the excitons migrating towards the nanobubble. Finally, from 5 ns to 20 ns, as the central population decays to the baseline, we revert to single-Gaussian fitting for the remaining population now trapped in the nanobubble, with the Gaussian center representing the position of the nanobubble. The blue line in the figure illustrates the motion of excitons as determined from this Gaussian fitting.